\documentclass[fleqn,twoside]{article}
\usepackage{espcrc2}

\def\alt{\raise0.3ex\hbox{$\;<$\kern-0.75em\raise-1.1ex\hbox{$\sim\;$}}}
\def\agt{\raise0.3ex\hbox{$\;>$\kern-0.75em\raise-1.1ex\hbox{$\sim\;$}}}
\def\d{{\rm d}}
\newcommand{\be}{\begin{equation}}
\newcommand{\ee}{\end{equation}}
\newcommand{\bea}{\begin{eqnarray}}
\newcommand{\eea}{\end{eqnarray}}
\newcommand{\vv}{\,\,\, ,}
\newcommand{\pp}{\,\,\, .}

\usepackage{graphicx}
\usepackage[figuresright]{rotating}

\newcommand{\AmS}{{\protect\the\textfont2
  A\kern-.1667em\lower.5ex\hbox{M}\kern-.125emS}}

\title{The Relevance of Mediterranean Neutrino Telescope Sites on Earth-skimming tau neutrino
detection}

\author{G. Miele\address[Naples]{Dipartimento di Scienze Fisiche, Universit\`{a} di Napoli
$Federico$ $II$ and
        INFN Sezione di Napoli,\\ Complesso
Universitario di Monte S.\ Angelo, Via Cinthia, I-80126 Napoli,
Italy.}\thanks{electronic address: miele@na.infn.it}}
\begin{document}

\begin{abstract}
A study of the UHE $\nu_\tau$ detection performances of a km$^3$
Neutrino Telescope sitting at the three proposed sites for
\verb"ANTARES", \verb"NEMO" and \verb"NESTOR" in the Mediterranean
sea is here performed. In particular, it is analyzed the effect of
the underwater surface profile on the total amount of yearly
expected $\tau$ crossing the fiducial volume in the limit of full
detection efficiency and energy resolution. \vspace{1pc}
\end{abstract}

\maketitle

\section{Introduction}

Neutrinos are one of the main components of the cosmic radiation
in the ultra-high energy (UHE) regime.  Although their fluxes are
uncertain and depend on the production mechanism, their detection
can provide information on the sources and origin of the UHE
cosmic rays. For example, UHE neutrinos can be produced via
$\pi$-photo\-production by strongly accelerated hadrons in
presence of a background electromagnetic field.  This scenario is
expected to occur in extreme astrophysical environments like the
jets of active galactic nuclei, radio galaxies and gamma ray burst
sources as well as in the propagation of UHE nucleons scattering
off the cosmic background radiation (known as {\it cosmogenic
neutrinos} \cite{[1],[2]}).

From the experimental point of view, after the first pioneering
and successful achievements neutrino astronomy in the high energy
regime \cite{[3],[4],[5],[6],[7]} is a rapidly developing field,
with a new generation of neutrino telescopes on the way, as in the
deep water of the Mediterranean sea, namely \verb"ANTARES"
\cite{[16]}, \verb"NESTOR" \cite{[17]} and \verb"NEMO"
\cite{[18]}. In the future they could lead to the construction of
a km$^3$ telescope as pursued by the \verb"KM3NeT" project
\cite{[19],[20]}.

Although NTs were originally thought as $\nu_{\mu}$ detectors,
their capability as $\nu_{\tau}$ detectors has become a hot topic
\cite{[24],[25],Anchordoqui:2005is,Yoshida:2003js}, in view of the
fact that flavor neutrino oscillations lead to nearly equal
astrophysical fluxes for the three neutrino flavors. Despite the
different behavior of the produced tau leptons with respect to
muons in terms of energy loss and decay length, both $\nu_\mu$ and
$\nu_{\tau}$ detection are sensitive to the matter distribution
near the NT site \cite{Cuoco:2006qd}. Thus, a computation of the
event detection rate of a km$^3$ telescope requires a careful
analysis of the surroundings of the proposed site. The importance
of the elevation profile of the Earth surface around the detector
was already found of some relevance in Ref.\
\cite{Miele:2005bt,Miele:2004ze}, where the present author and
others calculated the aperture of the Pierre Auger Observatory
\cite{Auger,Abraham:2004dt} for Earth-skimming UHE $\nu_{\tau}$'s.

In this paper it is computed the effective aperture for $\nu_\tau$
detection of a km$^3$ NT in the Mediterranean sea placed at any of
the three locations proposed by the \verb"ANTARES", \verb"NEMO"
and \verb"NESTOR" collaborations. A more complete treatment of the
subject also including the $\nu_\mu$ contribution is contained in
Ref. \cite{Cuoco:2006qd}.

\begin{figure}[t]
\begin{center}
\includegraphics[width=.40\textwidth]{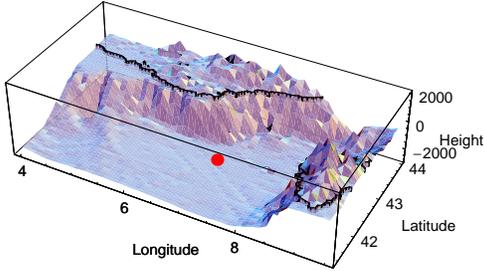}
\vspace{0pc} \caption{The surface profile of the area near the
\texttt{ANTARES} site (red spot) at 42$^\circ$ 30' N, 07$^\circ$
00' E. The black curve represents the coast line. The sea plateau
depth in the simulation is assumed to be 2685 m.} \label{Antares}
\end{center}
\end{figure}

\begin{figure}[t]
\begin{center}
\includegraphics[width=.40\textwidth]{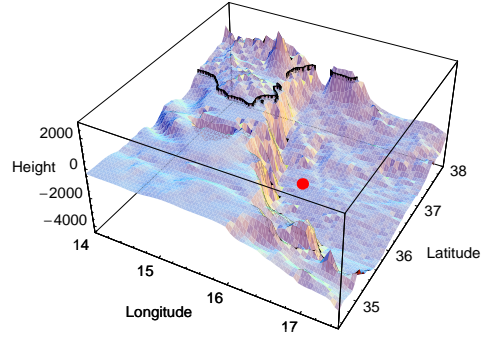}
\vspace{-3pc} \caption{The surface profile of the area near the
\texttt{NEMO} site (red spot) at 36$^\circ$ 21' N, 16$^\circ$ 10'
E. The black curve represents the coast line. The sea plateau
depth used in the simulation is 3424 m.} \label{Nemo}
\end{center}
\end{figure}

\begin{figure}[t]
\begin{center}
\includegraphics[width=.40\textwidth]{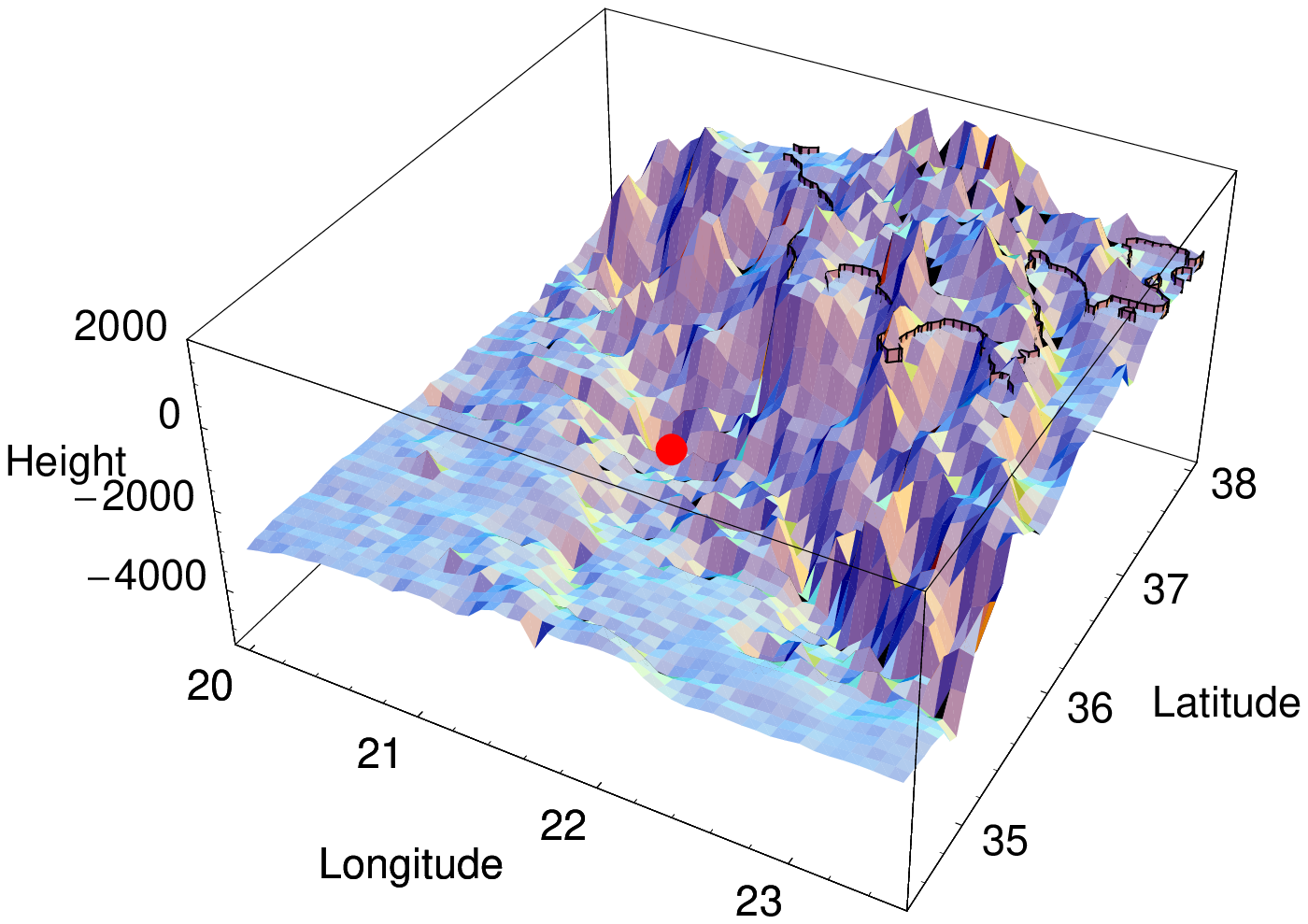}
\vspace{-2pc} \caption{The surface profile of the area near the
\texttt{NESTOR} site (red spot) at 36$^\circ$ 21' N, 21$^\circ$
21' E. The black curve represents the coast line. The sea plateau
depth in the simulation is assumed to be 4166 m.}
\vspace{-1pc}\label{Nestor}
\end{center}
\end{figure}

A detailed DEM of the under-water Earth surface is available from
the Global Relief Data survey (ETOPO2) \cite{ETOPO2}, a grid of
altimetry measurements with a vertical resolution of 1 m averaged
over cells of 2 minutes of latitude and longitude. In Figures
\ref{Antares}, \ref{Nemo} and \ref{Nestor} are shown the 3D maps
of the areas around the three NT sites. The black curve represents
the coast line, whereas the red spot stands for the location of
the apparatus. By following the same approach developed in
\cite{Miele:2005bt}, this DEM is used to produce a realistic and
statistically significant sample of $\nu_\tau$/$\tau$ tracks
crossing the fiducial volume of the NT that are then used to
evaluate the effective aperture of each detector.

\begin{figure}[!h]
\begin{center}
\includegraphics[width=.40\textwidth]{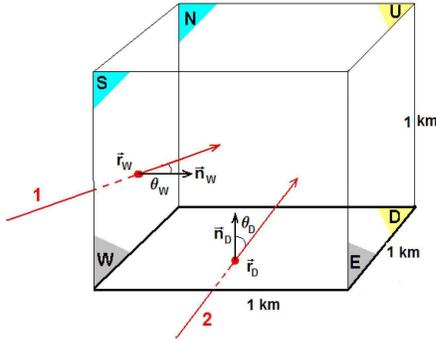}
\vspace{-3pc} \caption{The angle definition and the fiducial
volume of a km$^3$ NT.} \label{kmcube}
\end{center}
\end{figure}
The structure of the paper is as follows. In Section~\ref{ev_rate}
the formalism and definitions used in the analysis are introduced,
and the aperture for a NT is defined. The results for $\nu_\tau$
induced events are reported and discussed in Section~\ref{nu_tau}
for various incoming neutrino fluxes. Finally, the conclusions are
reported in Section~\ref{concl}.

\section{The effective aperture of a NT}\label{ev_rate}

Let us define the km$^3$ NT {\it fiducial} volume as that bounded
by the six lateral surfaces $\Sigma_a$ (the subindex $a$=W, E, N,
S, U and D labels each surface through its orientation: West,
East, North, South, Up, and Down), and indicate with $\Omega_a
\equiv (\theta_a, \phi_a)$ the generic direction of a track
entering the surface $\Sigma_a$. The scheme of the NT fiducial
volume and two examples of incoming tracks are shown in Fig.\
\ref{kmcube}.

Let $\d \Phi_\nu/(\d E_\nu \, \d\Omega_a)$ be the differential
flux of UHE $\nu_\tau + \bar{\nu}_\tau$. The number per unit time
of $\tau$ leptons emerging from the Earth surface and entering the
NT through $\Sigma_a$ with energy $E_\tau$ is given by
\bea \left( \frac{\d N_\tau}{\d t} \right)_a = \int \d\Omega_a
\int \d S_a \int \d E_\nu \, \frac{\d\Phi_\nu(E_\nu,
\Omega_a)}{\d E_\nu\,\d\Omega_a}\nonumber \\
\times\int \d E_\tau \cos\left(\theta_a\right)
 k_a^\tau(E_\nu,E_\tau;\vec{r}_a,\Omega_a).
\label{eq:1} \eea
This equation is the same as that in \cite{Miele:2005bt}, but for
full duty cycle and detection efficiency.  The kernel
$k_a^\tau(E_\nu\,,E_\tau\,;\vec{r}_a,\Omega_a)$ is the probability
that an incoming $\nu_\tau$ crossing the Earth, with energy
$E_\nu$ and direction $\Omega_a$, produces a $\tau$-lepton which
enters the NT fiducial volume through the lateral surface $\d S_a$
at the position $\vec{r}_a$ with energy $E_\tau$ (see Fig.\
\ref{kmcube} for the angle definition). If one splits the possible
events between those with track intersecting the {\it rock} and
the ones only crossing {\it water}, the kernel
$k_a^\tau(E_\nu\,,E_\tau\,;\vec{r}_a,\Omega_a)$ is given by the
sum of these two mutually exclusive contributions,
\begin{eqnarray}
k_a^\tau(E_\nu\,,E_\tau\,;\vec{r}_a,\Omega_a) =
k_a^{\tau,{r}}(E_\nu\,,E_\tau\,;\vec{r}_a,\Omega_a) \nonumber\\+
k_a^{\tau,{w}}(E_\nu\,,E_\tau\,;\vec{r}_a,\Omega_a) \pp
\label{kern-split}
\end{eqnarray}

For an isotropic flux one can rewrite Eq. (\ref{eq:1}), summing
over all the surfaces, as
\begin{eqnarray}
\frac{\d N_\tau^{(r,w)}}{\d t} = \int \d E_\nu \,
\,\frac{1}{4\pi}\,\frac{\d\Phi_\nu(E_\nu)}{\d E_\nu}
\,A^{\tau(r,w)}(E_\nu) \nonumber \\ = \sum_a \int \d E_\nu \,
\,\frac{1}{4\pi}\,\frac{\d\Phi_\nu(E_\nu)}{\d E_\nu}
\,A_a^{\tau(r,w)}(E_\nu) \vv \label{kernel1}
\end{eqnarray}
which defines the total aperture $A^{\tau(r,w)}(E_\nu)$, with
"$r$" and "$w$" denoting the {\it rock} and {\it water} kind of
events, respectively. The contribution of each surface to the
total aperture reads
\begin{eqnarray}
A_a^{\tau(r,w)}(E_\nu) = \int \d E_\tau \int \d\Omega_a \int  \d
S_a \, \cos\left(\theta_a\right) \nonumber\\ \times \,
k_a^{\tau,{(r,w)}}(E_\nu\,,E_\tau\,;\vec{r}_a,\Omega_a) \pp
\label{kernel2}
\end{eqnarray}

\section{The event rate for $\nu_\tau$ interactions}\label{nu_tau}

\begin{figure}[t]
\begin{center}
\includegraphics[width=.45\textwidth]{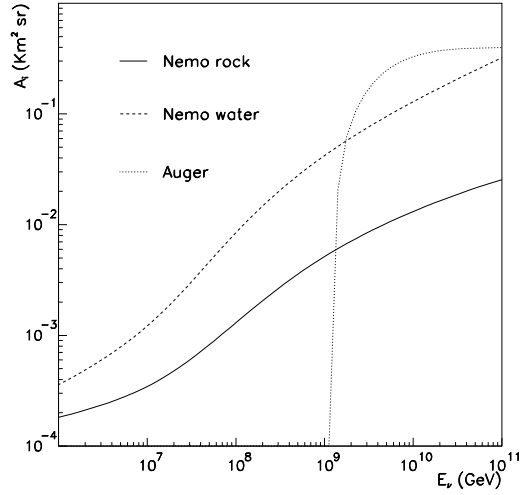}
\vspace{-2pc} \caption{The effective apertures
$A^{\tau(r)}(E_\nu)$ (solid line) and $A^{\tau(w)}(E_\nu)$ (dashed
line) defined in Eq.\ (\ref{kernel1}) versus neutrino energy for
\texttt{NEMO}. The dotted line corresponds to the same quantity
for the Auger Fluorescence Detector for Earth-skimming $\nu_\tau$
as in \cite{Miele:2005bt}.} \label{apertures}
\end{center}
\end{figure}
In Fig.\ \ref{apertures} are shown the apertures $A^{\tau(r,w)}$
for the \verb"NEMO" site together with the corresponding quantity
for the Pierre Auger Observatory Fluorescence Detector (FD)
calculated in \cite{Miele:2005bt}. Note that the Auger case is
only for Earth-skimming $\tau$'s, since down-going neutrino
induced events can be disentangled from ordinary cosmic rays only
for very inclined showers. Interestingly, the \verb"NEMO"-{\it
water} and Auger-FD apertures almost match at the FD threshold of
$10^{18}\,$eV, so that using both detectors results into a wide
energy range of sensitivity to $\nu_\tau$ fluxes.
\begin{figure*}[tb]
\begin{center}
\begin{tabular}{cc}
\includegraphics[width=.45\textwidth]{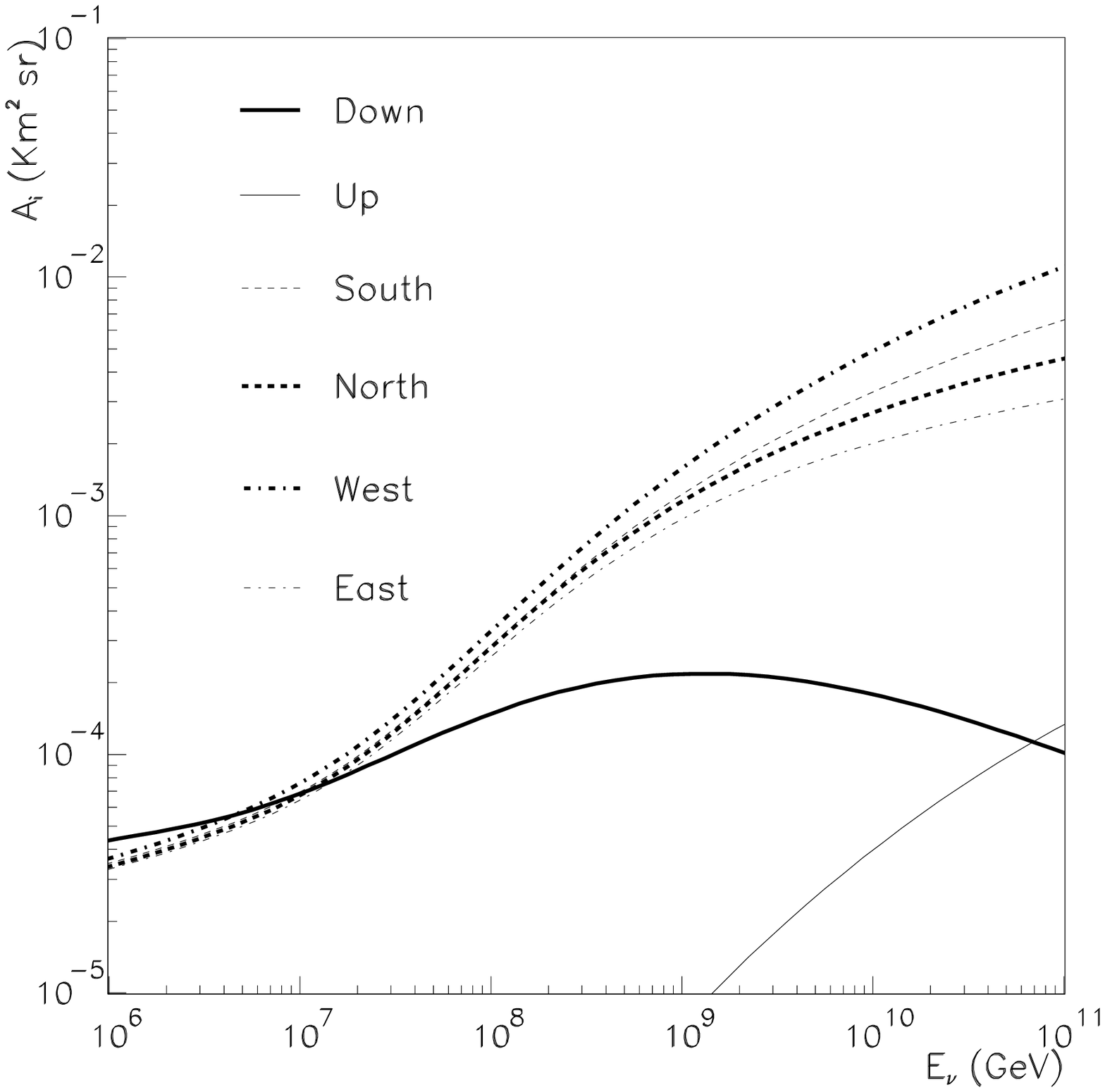} &
\includegraphics[width=.45\textwidth]{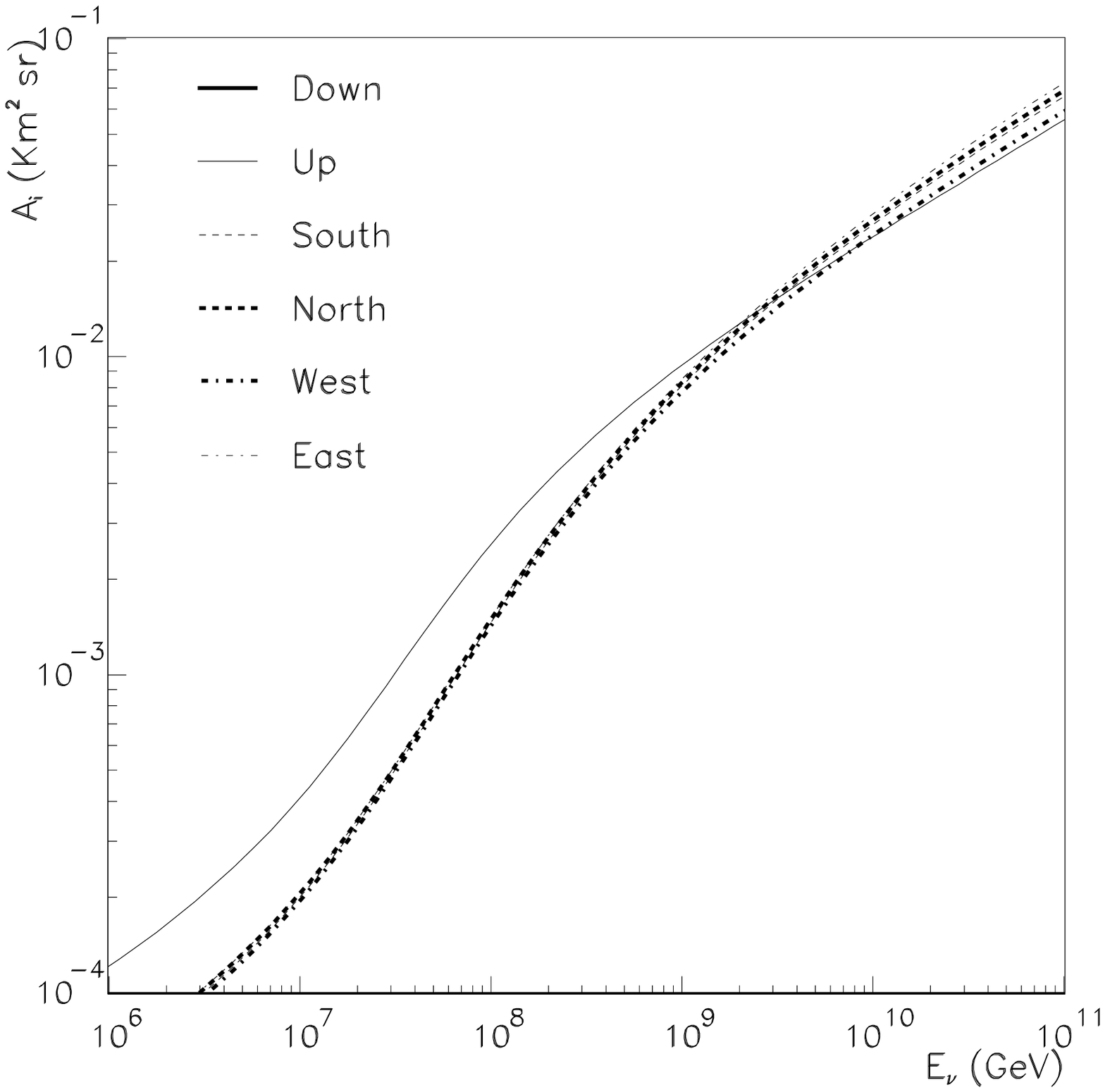}
\end{tabular}
\end{center}
\vspace{-3pc} \caption{The effective apertures
$A_a^{\tau(r,w)}(E_\nu)$ of Eq.\ (\ref{kernel2}) versus neutrino
energy for (left) {\it rock} events and (right) {\it water} events
for the \texttt{NEMO} site.} \label{apertsurf}
\end{figure*}

In Fig. \ref{apertsurf} we show the high energy behavior for each
surface contributing to the effective aperture. For {\it rock}
events there is a clear W-E asymmetry, easily understood in terms
of matter effects related to the particular morphology of the
\verb"NEMO" site (see Fig.\ \ref{Nemo}). A much smaller S-N
asymmetry is also present. For neutrino energies larger than
$10^7$ GeV the main contribution to the aperture
$A^{\tau(r)}(E_\nu)$ comes from the lateral surfaces, i.e.\ from
$\tau$ leptons emerging from the rock far from the NT basis and
crossing the fiducial volume almost horizontally. Instead, the
upper surface contribution is negligible due to the very small
fraction of events crossing the rock and entering the detector
from above. For {\it water} events the contribution to the
aperture from all surfaces is comparable (except for the lower one
which has no events), the upper one providing a slightly larger
contribution as the energy decreases. Indeed, events which would
cross the lateral surfaces should travel over a longer path in
water and this becomes more unlikely at lower energies due to the
shorter $\tau$ decay length.

\begin{figure}[t]
\vspace{-0.7pc}
\begin{center}
\includegraphics[width=.45\textwidth]{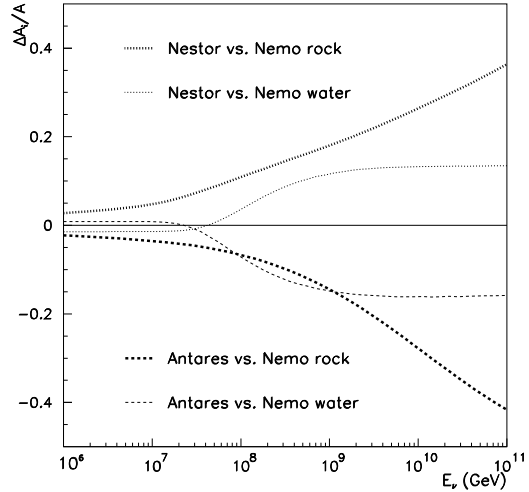}
\vspace{-2.5pc} \caption{A comparison of the effective NT
apertures with respect to the \texttt{NEMO} one is shown versus
the neutrino energy (see Eq. \ref{diffaper} for definitions).}
\label{comptelesc}
\end{center}
\end{figure}

In Fig. \ref{comptelesc} the detection performances of a km$^3$ NT
placed at one of the three sites in the Mediterranean sea are
compared. In particular the quantities
\begin{eqnarray}
\frac{\left[A^{\tau((r,w))}(\mbox{\texttt{NESTOR}})-A^{\tau((r,w))}(\mbox{\texttt{NEMO}})\right]}
{A^{\tau((r,w))}(\mbox{\texttt{NEMO}})}  \nonumber \\
\frac{\left[A^{\tau((r,w))}(\mbox{\texttt{ANTARES}})-A^{\tau((r,w))}(\mbox{\texttt{NEMO}})\right]}
{A^{\tau((r,w))}(\mbox{\texttt{NEMO}})} \label{diffaper}
\end{eqnarray}
are plotted versus the neutrino energy.

The \verb"NESTOR" site shows the highest values of the
$\tau$-aperture for both {\it rock} and {\it water}, due to its
larger depth and the particular matter distribution of the
surrounding area, while the lowest rates are obtained for
\verb"ANTARES". The aperture in the three sites can be quite
different at high energy but the net effect of this on the
expected number of events per year is not particularly significant
since the UHE neutrino flux drops rapidly with the energy.

Knowing the aperture of the NT at each site, one can compute the
expected $\tau$ event rate, once a neutrino flux is specified. In
Table \ref{table::events-WB} these rates are shown assuming a
GZK-WB flux. The enhancement effect due to the local matter
distribution is responsible for the N-S, W-E and NE-SW asymmetries
for the \verb"ANTARES", \verb"NEMO" and \verb"NESTOR" sites,
respectively, as expected from the matter profiles shown in Figs.\
\ref{Antares}, \ref{Nemo} and \ref{Nestor}. These matter effects,
for the specific UHE flux considered (GZK-WB), correspond to an
enhancement of {\it rock} events of the order of 20\% and a
screening factor for {\it water} events of the order of a few
percent. This is easily understood since for {\it water} events
the U surface gives a significant contribution which is
essentially unaffected by matter distribution.

It is important to emphasize that the role of matter effects
depends critically upon the energy spectrum of the UHE neutrino
flux. For more energetic neutrino fluxes the enhancement factor is
expected to be more significant (see the energy behavior of
$A_a^{\tau(r)}(E_\nu)$ in Fig \ref{apertsurf}).

In Table \ref{table::events} the rate of {\it rock/water} $\tau$
events are computed for the three different km$^3$ NT sites using
several UHE neutrino fluxes as already considered in
\cite{Miele:2005bt,Aramo:2004pr} and described in
\cite{Waxman:1998yy}-\cite{Bhattacharjee:1998qc}. As can be seen
from Table \ref{table::events}, the enhancement factors due to
matter effects on {\it rock} events can be as large as 30\%,
whereas the difference in the rates of {\it water} events for a
fixed neutrino flux is mainly due to the different depth of the
three sites.

\begin{table}[t]
{\footnotesize \centering
\begin{tabular}{|c|c|c|c|}
\hline Surf.
 & \verb"ANTARES" & \verb"NEMO" & \verb"NESTOR"\\
\hline
D & 0.0086/0 & 0.0086/0 & 0.0085/0 \\
U & 0/0.1741 & 0.0002/0.2191 & 0.0003/0.2595 \\
S & 0.0219/0.1635 & 0.0290/0.1799 & 0.0287/0.2014 \\
N &  0.0275/0.1573& 0.0268/0.1847 & 0.0359/0.1938 \\
W & 0.0247/0.1616 & 0.0371/0.1715 & 0.0303/0.2020 \\
E & 0.0240/0.1622 & 0.0228/0.1900 & 0.0414/0.1858 \\
\hline Total & 0.107/0.819 & 0.124/0.945 & 0.145/1.043\\
\hline
\end{tabular}\caption{Estimated rate per year of {\it rock/water}
$\tau$ events at the three km$^3$ NT sites for a GZK-WB flux. The
contribution of each detector surface to the total number of
events is also reported.} \label{table::events-WB}}
\end{table}

An interesting feature is the dependence of the event rate upon
the shape of the NT detector for a fixed total volume of 1 km$^3$,
a property that might be relevant for the eventual design of the
detector. Consider for example a km$^3$ NT placed at the
\verb"NEMO" site with the shape of a parallelepiped rather than a
cube, where in particular the E and W surfaces are enlarged by a
factor 3 in the horizontal dimension, the N and S surfaces being
reduced by the same factor, keeping the height of towers still of
1 km. In this case the expected rate of {\it rock} events per year
is enhanced by almost a factor 2, from 0.12 to $0.21$ for the
GZK-WB flux, while this enhancement could be even larger for
neutrino fluxes with a larger high energy component. Moreover, the
expected rate of {\it water} events increases as well by a factor
of the order of 50\%, from $0.945$ up to $1.425$ per year. Similar
exercises can be also performed for the \verb"ANTARES" and
\verb"NESTOR" sites. Of course, a further possibility which might
favor UHE-$\tau$ detection consists in increasing the effective
volume of the detector keeping unchanged the 1 km height and the
number of towers of photomultipliers but adopting a larger
spacing. As an example, for a factor four larger volume with a
doubling of the tower spacing both the {\it rock} and {\it water}
$\tau$ events would increase by almost a factor two, but obviously
at the expense of the energy threshold and the quality of the
event reconstruction for ``low-energy'' (TeV) neutrinos.  For a
detector aiming at the exploration of the range above the PeV,
this is a less severe problem.
\begin{table}[t]
\vspace{0.5pc} {\footnotesize \centering
\begin{tabular}{|c|c|c|c|}
\hline $\nu$-fluxes
 & \verb"ANTARES" & \verb"NEMO" & \verb"NESTOR" \\
\hline
GZK-WB& 0.107/0.819 & 0.124/0.945 & 0.145/1.043 \\
GZK-L & 0.110/1.106 & 0.141/1.308 & 0.175/1.471 \\
GZK-H & 0.245/2.284 & 0.335/3.348 & 0.423/3.777 \\
NH    & 1.029/8.920 & 1.248/10.37 & 1.488/11.51 \\
TD    & 0.837/5.181 & 0.957/5.888 & 1.087/6.436 \\
\hline
\end{tabular}
\caption{Yearly rate of {\it rock/water} $\tau$ events at the
three km$^3$ NT sites for different UHE neutrino fluxes.  GZK-H is
for an initial proton flux $\propto 1/E$, assuming that the EGRET
flux is entirely due to $\pi$- photoproduction.  GZK-L shows the
neutrino flux when the associated photons contribute only up to
20\% in the EGRET flux. GZK-WB stands for an initial proton flux
$\propto 1/E^2$ \cite{Waxman:1998yy}--\cite{Semikoz:2003wv}. The
other two neutrino fluxes correspond to more exotic UHECR models.
NH represents the neutrino flux prediction in a model with new
hadrons \cite{Kachelriess:2003yy}, whereas TD is the neutrino flux
for a topological defect model \cite{Bhattacharjee:1998qc}.}
\label{table::events}}
\end{table}

The fact that the event rate depends upon the total surface of the
detector is a peculiar feature of a NT, quite differently from
what expected at the Auger observatory. Actually, in this case
observed showers are generally initiated not very far from the
detector compared to its dimensions so that the shape of the
detector (i.e., the position on the border where the FDs are
placed) is not as important as its ``volume'' (controlled by the
area enclosed by the FDs). In fact, in order to produce a $\tau$
emerging from the Earth with enough energy to generate a shower
detectable by the Auger FDs (at least 1 EeV$=10^{18}$ eV), the
energy of the neutrino should be larger than $10^{18}$ eV, taking
into account the $\tau$ energy loss in the rock.  But the decay
length of such a UHE $\tau$ is $l_{\rm decay} \simeq 50\,$km
$\times (E_\tau/{\rm EeV})$, to be compared with the dimensions of
the Auger fiducial volume, $\sim 50\times 60\times10$ km$^3$.
Conversely, a neutrino telescope can detect tau's or muons which
are produced very far from the detector by a neutrino
charged-current interaction, from distances comparable to the
charged lepton range at that particular energy
\cite{Yoshida:2003js}. Indeed, the $\tau$ range in water is of the
order of several kilometers: from the value of $\beta_{\tau}=0.71
\times 10^{-6}$ cm$^2$ g$^{-1}$ one obtains an attenuation length
$1/(\beta_{\tau}\varrho_{w}) \simeq 15$ km, while for muons (see
\cite{Cuoco:2006qd}) the range is approximately eight times
smaller, of the order of 2 km. In other words, the effective
volume of a NT of the kind discussed so far can be much larger
than 1 km$^3$, thus maximizing the detector area might greatly
improve the detection rate.

Of course, one should not forget that the design of a NT also
depends strongly upon more detailed experimental considerations.
Shapes which are not very compact or a detector with very sparse
instrumentation have worse performances in the reconstruction of
track properties as well as in signal-background separation,
though this is mainly problematic at energies lower than 100 TeV,
in the atmospheric neutrino energy range. In any case, the present
analysis suggests that the choice of the detector shape could be
an important feature in orienting the target of a NT investigation
towards either atmospheric or extra-atmospheric neutrino physics.
In this respect, the possibility to modify this parameter quite
easily for a NT water detector offers a great advantage with
respect to an under-ice detector.

\begin{figure*}[!tb]
\begin{center}
\begin{tabular}{cc}
\includegraphics[width=0.48\textwidth]{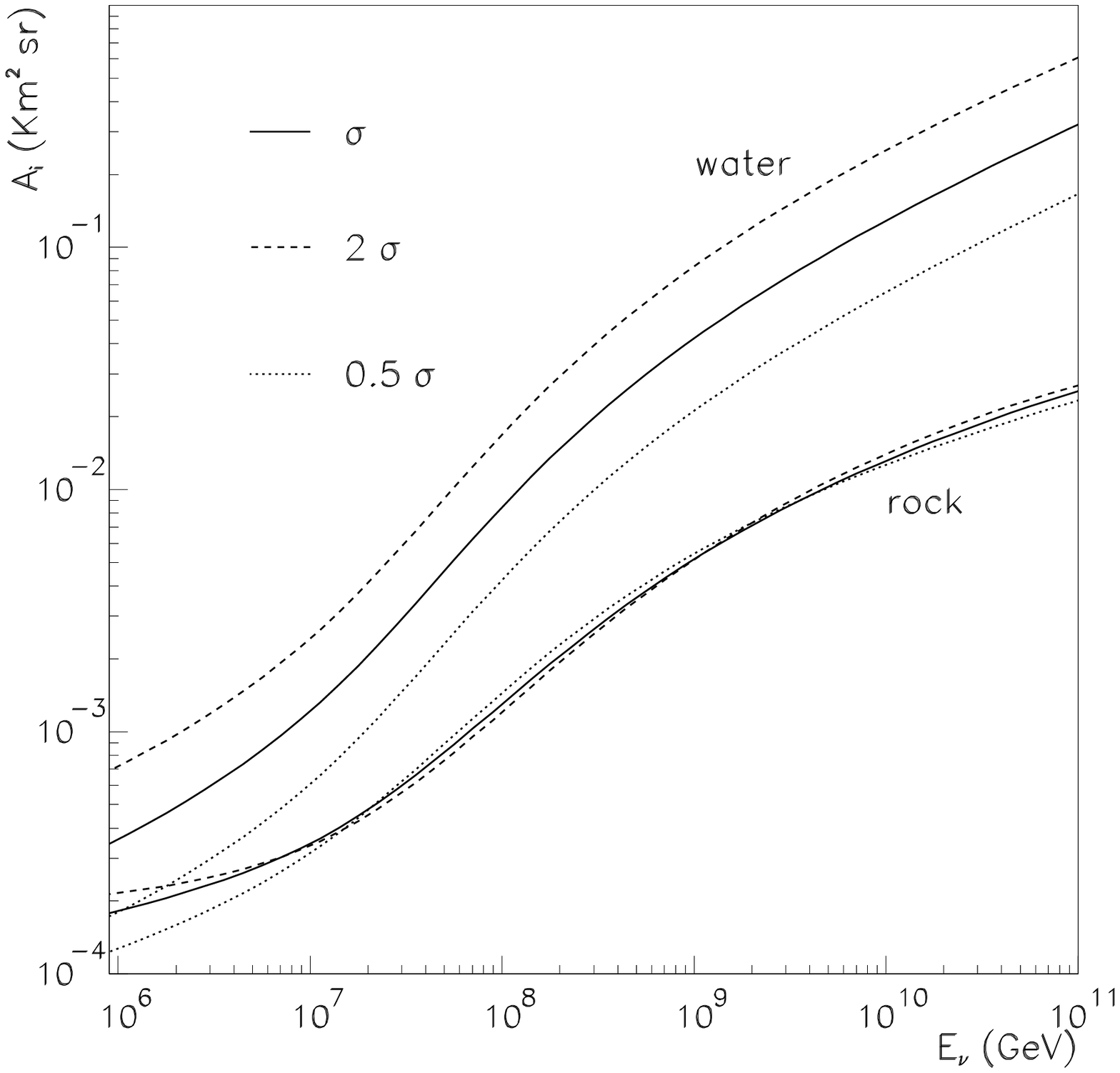} &
\includegraphics[width=0.46\textwidth]{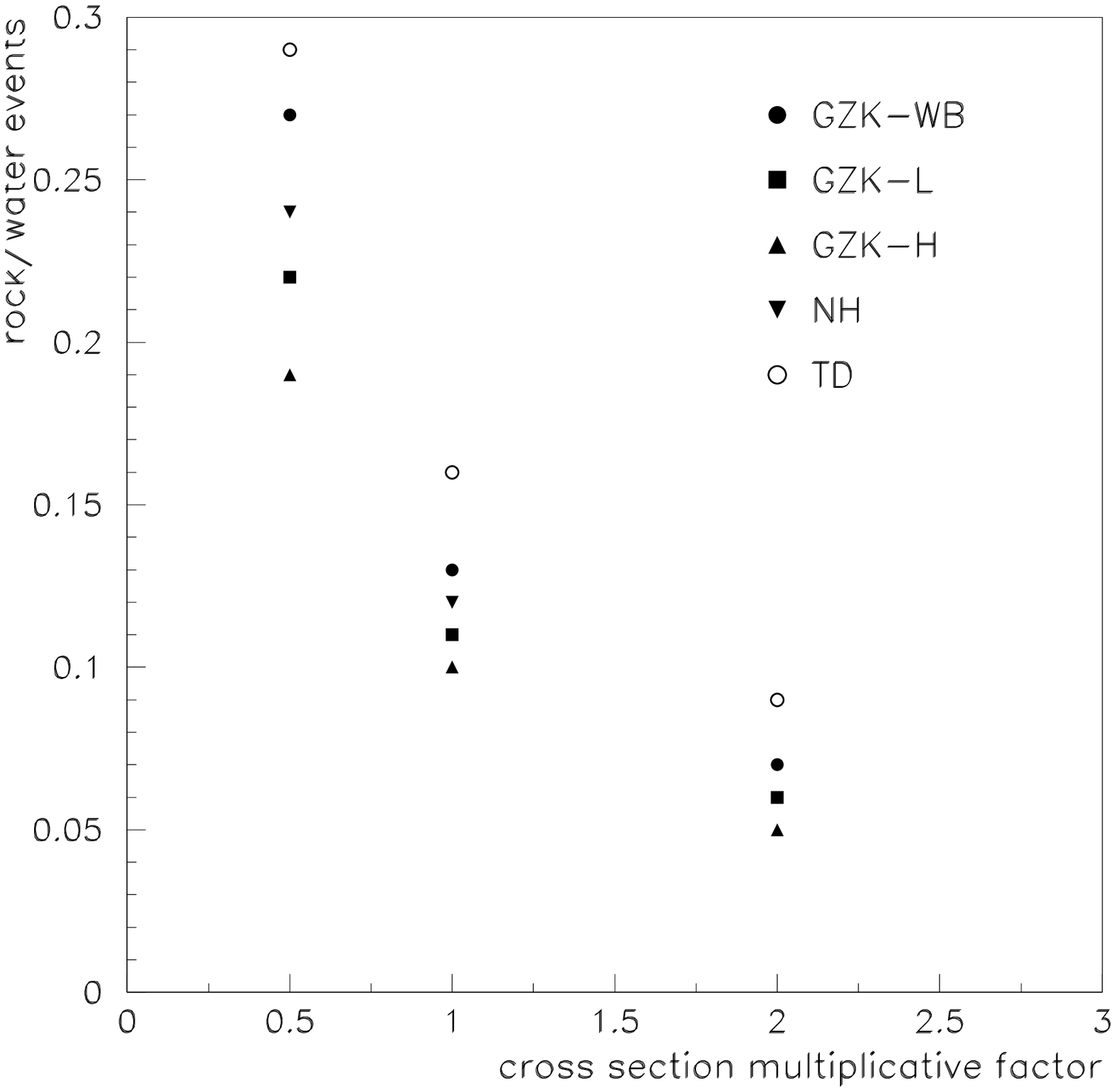}
\end{tabular}
\end{center}
\vspace{-3pc} \caption{(Left) The effective apertures
$A^{\tau(r,w)}(E_\nu)$ for the \texttt{NEMO} site for a
$\nu_\tau$--nucleon cross section multiplied by a factor 0.5, 1
and 2 with respect to the standard result $\sigma$. (Right) Ratios
of the number of events in {\it rock/water} when the cross section
is rescaled by the factor shown on the $x$ axis, for several
incoming UHE neutrino fluxes.} \label{AperEventsig}
\end{figure*}

One of the main motivations for studying UHE neutrinos is that
they provide a possibility to explore a range of energies for
scattering processes which is still untested (maybe impossible to
test) by particle accelerators. In this respect, measuring the
neutrino--nucleon cross sections at high energies could have a
large impact on constraining or discovering new physics beyond the
standard model (see e.g.\
\cite{Anchordoqui:2005is,Anchordoqui:2005gj}).  While a
measurement of the event energy spectrum cannot remove in general
a degeneracy between the neutrino cross section and the incoming
neutrino flux, a neutrino telescope could offer the interesting
capability of disentangling these two factors because of the role
of matter effects. Indeed, provided that enough statistics is
collected and the detector has a good zenith angle resolution, the
flux dependence can be subtracted off by measuring the ratio of
the event rates coming from different directions
\cite{Kusenko:2001gj,Hooper:2002yq,Hussain:2006wg}. In the left
panel of Fig.\ \ref{AperEventsig} we show how the \verb"NEMO"
effective apertures for $\tau$ {\it rock} and {\it water} events
change if the neutrino-nucleon cross section is half or twice the
standard model result while in the right panel we display the
ratio of {\it water/rock} $\tau$ event rates for several adopted
fluxes. One can see that this ratio is quite sensitive to the
value of the cross section. In particular, the number of {\it
rock} events is essentially unaffected while the {\it water} event
rate increases almost linearly with the cross section. Clearly,
since the statistical error on the ratio would be dominated by the
rare {\it rock} events, an experiment which aims at exploiting
this effect should maximize the acceptance for almost horizontal
events. It is important noticing that the present results do not
take into account any detailed experimental setup and up to this
point the $\nu_\mu$ contribution is not yet considered.
Nevertheless, since both the incoming neutrino flux and cross
section on nucleons are expected to be flavor independent the
possibility of determining both these quantities at a NT seems an
interesting perspective.

\section{Conclusions}\label{concl}

Ultra-high energy neutrinos represent one of the main targets for
several experiments which adopt a variety of detection techniques.
Among these, the optical Cherenkov neutrino telescopes, deployed
under water or ice, look for the tracks of charged leptons
produced by the high energy neutrinos that reach the Earth. In
this paper, it has been presented a new study of the performance
of a km$^3$ neutrino telescope to be located in any of the three
sites in the Mediterranean sea proposed by the \texttt{ANTARES},
\texttt{NEMO}, and \texttt{NESTOR} collaborations. In particular
the details of the under-water surface profile of each of the
three sites, using the data from a Digital Elevation Map, have
been considered. By generating a realistic and statistically
significant sample of $\nu_\tau$/$\tau$ tracks crossing the
fiducial volume of the km$^3$ neutrino telescope, the effective
aperture to UHE $\nu_\tau$ neutrinos have been calculated, as well
as the expected number of events for different UHE neutrino
fluxes, for both cases where the charged lepton is produced in
rock and water near the telescope site.

Although  in the present paper the detector-specific features,
like partial detection efficiency and finite energy resolution,
have not included, the obtained results show that the impact of
the site geography (or matter effects) on observables such as the
total event rate or asymmetries in the event direction can be
important.

These matter effects can be enhanced by a suitable choice of the
geometry of the telescope with a fixed volume. We have also
briefly discussed the possibility of obtaining information on both
the neutrino flux and the neutrino-nucleon cross section in the
UHE range by measuring the ratio of rock and water events.

\end{document}